\documentclass[twocolumn,showpacs]{revtex4}
\usepackage{graphicx}
\usepackage{dcolumn}

\begin{document}


\title{Anomalous behavior of the first excited 0$^{+}$ state
       in $N \approx Z$ nuclei}

\author{K. Kaneko$^1$, R. F. Casten$^2$, M. Hasegawa$^3$, T. Mizusaki$^4$, \\
Jing-ye Zhang$^{2,5}$, E. A. McCutchan$^2$, N. V. Zamfir$^2$, R.
Kr\"{u}cken$^6$}
\affiliation{
$^1$Department of Physics, Kyushu Sangyo University, Fukuoka 813-8503, Japan \\
$^2$WNSL, Yale University, New Haven, Connecticut 06520-8124, USA\\
$^3$Laboratory of Physics, Fukuoka Dental College, Fukuoka
814-0193, Japan\\
$^4$Institute of Natural Sciences, Senshu University, Kawasaki,
Kanagawa, 214-8580, Japan\\
$^5$Department of Physics and Astronomy, University of Tennessee,
Knoxville, Tennessee 37996, USA\\
$^6$Physik Department E12, Technische Universi\"{a}t M\"{u}nchen,
85748 Garching, Germany }

\date{\today}

\begin{abstract}
A study of the energies of the first excited $0^+$ states in all
even-even $Z$ $\geq$ 8 nuclei reveals an anomalous behavior in
some nuclei with $N$ = $Z$, $Z$ $\pm$ 2.  We analyze these
irregularities in the framework of the shell model. It is shown
that proton-neutron correlations play an important role in this
phenomenon.

\end{abstract}


\maketitle A topic of current interest and varying interpretations
is the nature of excited $J^{\pi}$ = $0^+$ states in nuclei.
Traditionally, the lowest excited $0^+$ states in even-even
deformed nuclei have been interpreted as ``$\beta$''
vibrations\cite{bandm}. In recent years, various papers
\cite{randb,burke,kumar,gunther,garrett} have discussed the lowest
$K$ = $0^+$ excitation as a collective excitation built on the
$\gamma$ vibration.  The observation of numerous excited $0^+$
states in $^{158}$Gd\cite{shelly} prompted several theoretical
interpretations.  These include their description as two-phonon
octupole in character \cite{victor} or as quasi-particle
excitations based on the projected shell model\cite{sun}. In light
nuclei, descriptions in terms of pairing vibrations,
multi-particle - multi-hole intruder states and isobaric analog
states have been discussed for decades. Nevertheless, to date, a
complete understanding of the origin of excited $0^+$ states
remains elusive.

The purpose of this Rapid Communication is twofold. First, we will
show a striking, and heretofore unrecognized, anomaly in $0^+_2$
energies that occurs in certain (but not all) light nuclei with
$N$ = $Z$, $N$ = $Z$ $\pm$ 2. Secondly, we will present shell
model calculations that show the significant role of the T = 0
part of the residual proton-neutron interaction in these
anomalies.

With the high current interest in $N$ $\sim$ $Z$ nuclei and the
likelihood that new examples of such nuclei will be studied in
greater detail with exotic beam experiments in upcoming years, the
discovery of new phenomena in such nuclei takes on heightened
interest.  Looking over the entire nuclear chart, the trend of
first excited $0^+$ energies exhibits an interesting behavior. To
compare nuclei over a wide range of structures, we normalize the
energy of the $0_2^+$ state by the energy of the $2_1^+$ state,
defining $R_{0/2}$ $\equiv$ $E$($0_2^+$)/$E$($2_1^+$). This
$R_{0/2}$ ratio is plotted as a function of the energy of the
first $2_1^+$ state in Fig. 1, for all even-even nuclei with $Z$
$\geq$ 8.  For the majority of nuclei, the $R_{0/2}$ ratios follow
a compact trajectory.  There are, however, obvious deviations from
this trajectory: 7 points which clearly stand out above the main
trajectory.

\begin{figure}[t]
\includegraphics[width=8cm,height=7cm]{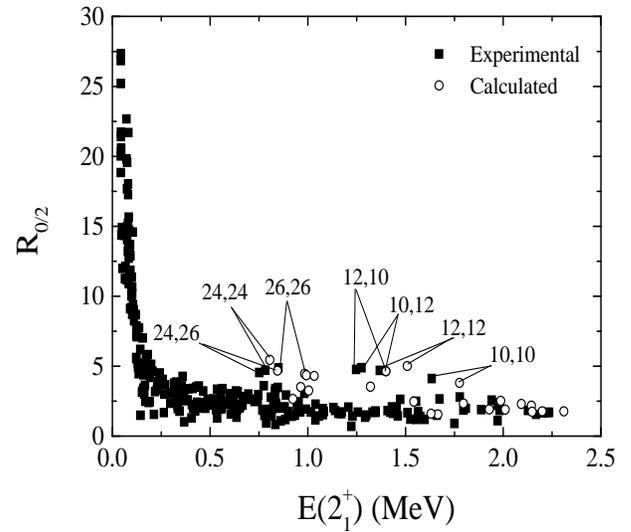}
\caption{The energy ratio $R_{0/2}$ $\equiv$
$E$($0_2^+$)/$E$($2_1^+$) as a function of the energy of the first
excited $2_1^+$ state for all even-even nuclei with $Z$ $\geq$ 8.
The points above the main trajectory are labelled by their ($Z,N$)
values.  The solid squares denote the experimental values and the
open circles denote the calculated values for some of these nuclei
(see text). The three unlabelled, calculated points that lie above
the trajectory near E($2_1^+$) $\sim$ 1.0 and 1.3 MeV correspond
to $^{44,46,48}$Ti.}
\label{fig1}
\end{figure}

Interestingly, the deviations in Fig. 1 comprise a set of nuclei
with $N$ = $Z$ and $N$ = $Z$ $\pm$ 2, specifically $^{20,22}$Ne,
$^{22,24}$Mg, $^{48,50}$Cr and $^{52}$Fe with $Z$ and/or $N$ = 10,
12, 24, and 26.  An important aspect of the anomaly is that not
all nuclei with $N$ = $Z$, $Z$ $\pm$ 2 exhibit it. Specifically,
nuclei with $N$ or $Z$ magic or with $N$ or $Z$ = 14, 16, 18 or 22
lie within the main trajectory. The phenomenon occurs when, in the
most simple view of shell filling, there is an open 1$d_{5/2}$
shell or an open 1$f_{7/2}$ shell, except for $N$,$Z$ = 22. Any
interpretation of this phenomenon must account not only for its
existence, but also its locus and must explain why it is not
universal in $N$ = $Z$ and $N$ = $Z$ $\pm$ 2 light nuclei.

Inspection of the individual excitation energies $E$($0_2^+$) and
$E$($2_1^+$) shows that the ratio $R_{0/2}$ is large for these
nuclei due to the combination of a large value of $E$($0_2^+$) and
a small value of $E$($2_1^+$) compared with neighbors in the
corresponding mass region. Since small $2_1^+$ excitation energies
are in general associated with soft or deformed nuclei, we
investigate the connection between $R_{0/2}$ and deformation in
Fig. 2, focusing on the region, 8 $\geq$$Z$$\geq$ 30, where the
anomalies are located. As a measure of the structure, we use the
energy ratio $R_{4/2}$ = E($4_1^+$)/E($2_1^+$), where $R_{4/2}$ =
2.0 for spherical nuclei, 3.33 for axially deformed nuclei. As the
low $2_1^+$ energy suggests, each of the anomalous nuclei have
$R_{4/2}$ $>$ 2.4. However, from Fig.2, one can also see that
there are several other nuclei which have similar structure but do
not exhibit an anomalous $R_{0/2}$. This observation reiterates
that the anomaly results from a combination of a low $2_1^+$
excitation energy and a high $0_2^+$ excitation energy.

\begin{figure}[t]
\includegraphics[width=8cm,height=7cm]{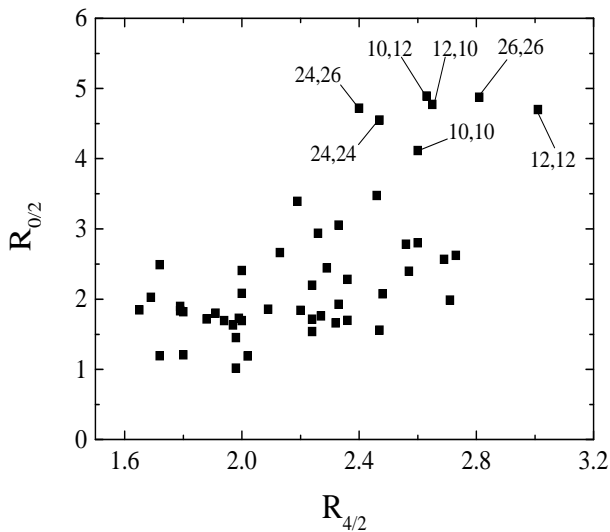}
\caption{The energy ratio $R_{0/2}$ as a function of $R_{4/2}$ for
$sd$ and $fp$ shell even-even nuclei with 8 $\leq$ $Z$ $\leq$ 30.
Those nuclei with anomalous $R_{0/2}$ values are labelled by their
(Z,N) values.}
\label{fig2}
\end{figure}
The remainder of this paper will focus on the origin of the high
lying $0_2^+$ states in these nuclei using shell model
calculations. Since the abnormalities in $R_{0/2}$ occur for
nuclei with $N$ $\approx$ $Z$, an appropriate aspect to consider
is $p-n$ interactions, which are strong for $N$ = $Z$ nuclei and
play a significant role in determining their structure
\cite{goodman}. Numerous studies of $p-n$ correlations
incorporating the competition between $T$ = 0 and $T$ = 1
components have investigated the energy spectrum as well as the
binding energy of proton rich nuclei. The large $p-n$ correlations
at $N$ $\approx$ $Z$ give rise to several interesting phenomena
such as singularities in the $p-n$ interaction
energy\cite{zhang,zamfir,kaneko}, alpha-like
correlations\cite{hasegawa}, the Wigner energy\cite{satula}, and
degenerate $T$ = 0 and $T$ = 1 lowest states in odd-odd, $N$ = $Z$
nuclei\cite{zeldes,brentano,vogel,macchavelli,satula2,kaneko2}. We
can expect that large $p-n$ correlations may also contribute to
the properties of the $0_2^+$ state in $N$ $\approx$ $Z$ nuclei.

In order to analyze the above behavior of the $0_2^+$ state in $N$
$\approx$ $Z$ nuclei, we perform full shell model calculations for
a wide range of nuclei. For the $sd$ shell, with model space
($1d_{5/2}$,$2s_{1/2}$,$1d_{3/2}$), we adopt the Wildenthal
interaction\cite{brown} to perform calculations for $^{20-26}$Ne,
$^{22-30}$Mg, $^{28-32}$Si, $^{32-34}$S, $^{36}$Ar. We note that,
in terms of energies, the results obtained for the $sd$ shell are
identical to those obtained in shell model calculations by
Brown\cite{website}.  For the $fp$ shell, with model space
($1f_{7/2}$,$2p_{3/2}$,$1f_{5/2}$,$2p_{1/2}$), we make use of the
KB3 interaction\cite{poves} for calculations on $^{44-50}$Ti,
$^{48-54}$Cr and $^{52-58}$Fe. Due to the enormous size of the
configuration space for $^{54-58}$Fe, we employed an extrapolation
method\cite{ex} for the shell model calculations of these nuclei.
The calculated values of $R_{0/2}$ are indicated by the open
circles in Fig. 1 and follow very well the experimental trend. In
particular, the large deviations from the simple trajectory for
the seven nuclei near $N$ $\approx$ $Z$ are well reproduced. The
calculations for the remaining nuclei lie within the main
trajectory, with the exception of 3 points corresponding to
$^{44-48}$Ti.  This peculiarity associated with the Ti nuclei will
be discussed below.

\begin{figure}[t]
\includegraphics[width=9cm,height=8cm]{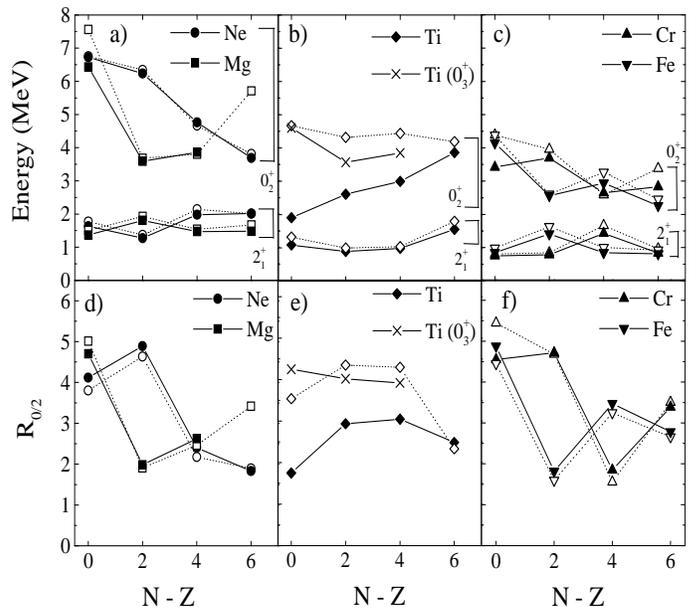}
  \caption{ Excitation energies of the first excited $0^+$ and $2^+$
states and as a function of $N$ - $Z$ for the (a) Ne and Mg (b) Ti
and (c) Cr and Fe isotopes. Ratio $R_{0/2}$ $\equiv$
$E$($0_2^+$)/$E$($2_1^+$) as a function of $N$ - $Z$ for the (d)Ne
and Mg (e) Ti and (f) Cr and Fe isotopes. The solid and open
symbols denote the experimental and calculated values,
respectively. The cross symbols in (b) and (e) correspond to the
second excited $0^+$ state in the Ti isotopes.}
  \label{fig3}
\end{figure}

A comparison with experimental $0_2^+$ and $2_1^+$ energies as a
function of $N$ - $Z$ for Ne and Mg is given in Fig. 3(a) and for
the Cr and Fe isotopes in Fig. 3(c). Included are all seven nuclei
previously highlighted, except for $^{22}$Mg which should be
identical to $^{22}$Ne assuming isospin invariance. For
$^{20,22}$Ne and $^{48,50}$Cr, i.e., with $N$ = $Z$, $Z$ + 2, the
$0_2^+$ energies are large while the $2_1^+$ energies are small.
As neutron number increases, the $0_2^+$ energy decreases and the
$2_1^+$ energy increases. In the Mg and Fe isotopes, the $0_2^+$
energy is large only for $^{24}$Mg and $^{52}$Fe ($N$ = $Z$) and
then decreases with neutron excess. The experimental trends of the
$0_2^+$ and $2_1^+$ energies are reproduced well by the
calculations.  Figures 3(d) and (f) show the corresponding
experimental and calculated $R_{0/2}$ ratios. As expected from the
discussion of the $0_2^+$ and $2_1^+$ energies, the ratios for
$^{20,22}$Ne and $^{48,50}$Cr are large and almost two times those
of the heavier Ne and Cr isotopes. In the Mg and Fe isotopes,
however, the ratios for $^{26}$Mg and $^{54}$Fe  ($N$ = $Z$ + 2,
with $N$ = 14 and $Z$ = 28) are small while those of $^{24}$Mg and
$^{52}$Fe ($N$ = $Z$) are large.  Again, the experimental behavior
is reproduced by the calculations.

While the anomaly for $N$=$Z$ and $N$=$Z$ $\pm$ 2 nuclei is
widespread in the open $1d_{5/2}$ and $1f_{7/2}$ shell nuclei, it
is conspicuously absent experimentally in the Ti isotopes.  In
contrast, in these same nuclei, the $0_2^+$ state and $R_{0/2}$
ratio is actually very low in energy as seen in Figs 3(b) and (e).
This result is not found in the calculations. They predict the
anomaly in $^{44,46}$Ti and even in $^{48}$Ti ($N$ = $Z$ +4). A
recent new effective interaction, GXPF1\cite{new}, yields similar
results for the Ti isotopes. The explanation of this lies in the
well-known presence of low lying $0^+$ intruder states in Ti
arising from 2p-2h proton and 2p-2h neutron excitations across the
$Z$,$N$ = 20 shell gap, which leads to a collective, low-lying
$0_2^+$ state. This state is clearly beyond the space of the
present calculations.  The reason that such a mode is not found in
Fe, for example, is simply that the excitation of 2 protons from
below Z = 20 in Fe ($Z$ = 26) would lead to a filled $f_{7/2}$
proton shell with the consequent decrease, rather than increase,
in the number of valence protons. Indeed, in other nuclei with a
filled $1f_{7/2}$ shell, (i.e., $^{54}$Fe) the anomaly is also not
found. The $0_3^+$ states in the Ti isotopes (crosses in Fig. 3)
likely correspond to an $fp$ shell excitation and their energies
are reproduced reasonably well by the calculations (open
diamonds).

While we have shown that shell model calculations can reproduce
the anomalous behavior of $R_{0/2}$, a more detailed analysis of
the calculations is required to understand how and why such good
agreement is obtained. We first consider the following
Hamiltonian:

\begin{equation}
H = H_{sp} + V_{int},
\end{equation}

\noindent where $H_{sp}$ is the single particle Hamiltonian and
$V_{int}$ is the realistic shell model interaction, such as the
Wildenthal interaction as used in the following analysis.  To
examine the roles of the $T$ = 0 and $T$ = 1 correlations in the
$0_2^+$ state, we separate the shell model interaction $V_{int}$
into two parts corresponding to the isoscalar, $V_{T=0}$, and
isovector, $V_{T=1}$, interactions as follows:

\begin{equation}
V_{int} = V_{T=0} + V_{T=1}
\end{equation}

To examine the contributions of the $T$ = 0 and $T$ = 1 components in
the shell model interaction, we can calculate their expectation
values in the eigenstates of the full Hamiltonian, Eq. (1). Since
the excitation energy depends strongly upon the competition
between the correlation energies of the excited state and the
ground state, we define the following correlation energy
difference:

\begin{equation}
\delta \langle O \rangle = \langle O \rangle_{ex} - \langle O
\rangle_{gr},
\end{equation}

\noindent where $\langle O \rangle_{gr}$ and $\langle O
\rangle_{ex}$ denote the expectation values of the operator $O$
for the ground and first excited $0^+$ states, respectively, and
$O$ is a physical operator such as the shell model interaction,
$V_{int}$ or the $T$ = 0, 1 interactions.
\begin{figure}[t]
\includegraphics[width=6cm,height=8cm]{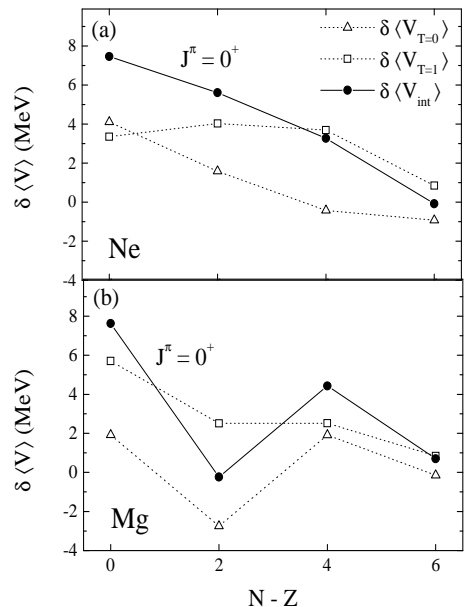}
  \caption{Correlation energy differences, $\langle$$V_{int}$$\rangle$ and
the corresponding $T$ = 0 and $T$ = 1 components for the (a) Ne and
(b) Mg isotopes.}
  \label{fig4}
\end{figure}

Figure 4 shows the correlation energy difference of $0^+$ states
as a function of $N$ - $Z$ in the Ne and Mg isotopes, including
the shell model interaction energy, $\langle$$V_{int}$$\rangle$,
and its decomposition into isoscalar, $\langle$$V_{T=0}$$\rangle$,
and isovector, $\langle$$V_{T=1}$$\rangle$, parts. From Fig. 4 we
see a correspondence between the ratio $R_{0/2}$ in Fig. 3(d) and
the correlation energy differences,
$\delta$$\langle$$V_{int}$$\rangle$. For each of the Ne and Mg
nuclei, with the exception of $^{20}$Ne, the T =1 component is
larger than the T = 0 component. In the Mg isotopes, Fig. 4(b),
the overall evolution of the correlation energy difference,
$\delta$$\langle$$V_{int}$$\rangle$, follows very closely the
behavior of the $T$ = 0 component.  Both the $R_{0/2}$ value and
correlation energy difference are large in $^{24}$Mg and decrease
suddenly in $^{26}$Mg.  This is attributed to the $T$ = 0
component, which displays a singular negative value for $N$ - $Z$
= 2, resulting in a sudden decrease in
$\delta$$\langle$$V_{int}$$\rangle$. For the Ne isotopes, shown in
Fig. 4(a), the $T$ = 1 component remains relatively constant
($\sim$3.5 MeV) for $N$ = 10 - 14 dropping to $\sim$ 1 MeV for $N$
= 16. Because of the nearly constant behavior of the $T$ = 1
component, the correlation energy differences in the Ne isotopes
follow the trend of the $T$ = 0 component, decreasing smoothly
with increasing neutron number. This behavior results in large
correlation energy differences in the Ne isotopes with $N$ = $Z$
and $N$ = $Z$ + 2, which will generate high excitation energies
for the $0_2^+$ state in these nuclei compared to isotopes with
larger $N$.  Thus, the behavior of correlation energy differences,
which strongly affects the ratio $R_{0/2}$, is attributed to the
combined effects of T=1 and T=0 components. The former contributes
the main amplitude, while the latter correlates with variations in
$R_{0/2}$.
\begin{figure}[b]
\includegraphics[width=6cm,height=8cm]{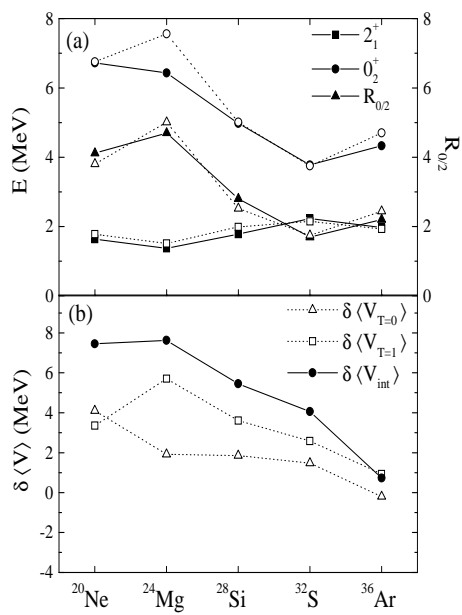}
  \caption{(a) Excitation energies of the $0_2^+$ and $2_1^+$ states and
the corresponding $R_{0/2}$ ratios for the $N$ = $Z$ $sd$-shell
nuclei. The solid and open symbols denote the experimental and
calculated values, respectively. (b) Correlation energy
differences $\langle$$V_{int}$$\rangle$ and the corresponding $T$ =
0 and $T$ = 1 components for the $N$ = $Z$ $sd$-shell nuclei.}
  \label{fig5}
\end{figure}

From the above discussion, we might expect large values of the
ratio $R_{0/2}$ for all nuclei with $N$ = $Z$.  However, as
previously mentioned, not all nuclei with $N$ = $Z$, $Z$ $\pm$ 2
are anomalous. The phenomenon occurs when there is an open
1$d_{5/2}$ shell or an open 1$f_{7/2}$ shell, except for N,Z = 22.
We now focus on the $N$ = $Z$ $sd$-shell nuclei and address why
only $^{20}$Ne and $^{24}$Mg exhibit large $R_{0/2}$ ratios.
Figure 5(a) illustrates the excitation energies of the $0_2^+$ and
$2_1^+$ states and the corresponding $R_{0/2}$ ratios.  Again, the
excitation energies are well reproduced by the shell model
calculations. Compared with neighboring $N$ = $Z$ nuclei,
$^{20}$Ne and $^{24}$Mg exhibit large excitation energies of the
$0_2^+$ states, small excitation energies of the $2_1^+$ states
and therefore large $R_{0/2}$ ratios. The correlation energy
differences, along with their $T$ = 0 and $T$ = 1 components, are
given in Fig. 5(b). The former are large for $^{20}$Ne and
$^{24}$Mg and decrease smoothly with increasing proton number. The
behavior of the $0_2^+$ states in Fig. 5(a) is similar to that of
the correlation energy differences,
$\delta$$\langle$$V_{int}$$\rangle$, in Fig. 5(b), with the
exception of $^{36}$Ar. A similar behavior as described above is
observed in the $N$ = $Z$ + 2 nuclei. For these nuclei, only
$^{22}$Ne displays a large $R_{0/2}$ ratio and with increasing
proton number, the ratio decreases rapidly.

In conclusion, we have investigated the first excited $0^+$ states
in light even-even nuclei. It is found that an anomaly occurs in
the ratio $R_{0/2}$ $\equiv$ $E$($0_2^+$)/$E$($2_1^+$) in some
even-even $N$ $\approx$ Z nuclei. The anomaly results from the
combination of large E($0^+_2$) energies and small E($2_1^+$)
energies. The latter are associated with a softness to deformation
in these nuclei. Shell model calculations reproduce well the
observed behavior. Concentrating on the Ne and Mg isotopes, the
anomalies of the $0_2^+$ states were analyzed using the
correlation energy differences and a decomposition of the shell
model interaction into $T$ = 0 and $T$ =1 components. From this
analysis, it is shown that the anomalous $R_{0/2}$ values can be
attributed to the joint effects of the T = 0 and T = 1 components:
the T=1 component contributes to the main amplitude while the T=0
component correlates with the variation of $R_{0/2}$.

We are grateful to B. A. Brown, R. Broglia, Larry Zamick and Igal
Talmi for inspiring discussions. This work was supported by U.S.
DOE Grant Nos. DE-FG02-91ER-40609 and DE-FG05-96ER-40983.


\end{document}